\documentclass[epj,final]{svjour}
\usepackage{amssymb}
\usepackage{amsmath}
\usepackage{graphics}
\usepackage{epsfig}

\begin{document}
\newcommand{\peeppi}{$p(e,e'p)\pi^0$~}
\newcommand{\peep}{$p(e,e'p)$~}
\newcommand{\pepi}{$p(e,e'p)\pi^0$~}
\newcommand{\D}{$\Delta (1232)$~}
\newcommand{\Ds}{${\rm D}_{13} (1900)$~}
\newcommand{\Ps}{${\rm P}_{11} (1440)$~}
\newcommand{\peeKL}{$p(\vec e,e'K \vec \Lambda)$~}
\newcommand{\snp}{S_{0+}^{*}}
\newcommand{\sem}{S_{1-}^{*}}
\newcommand{\sep}{S_{1+}^{*}}
\newcommand{\snpx}{S_{0+}}
\newcommand{\semx}{S_{1-}}
\newcommand{\sepx}{S_{1+}}
\newcommand{\enp}{E_{0+}}
\newcommand{\eem}{E_{1-}}
\newcommand{\eep}{E_{1+}}
\newcommand{\mem}{M_{1-}}
\newcommand{\mep}{M_{1+}}
\newcommand{\EMRpi}{\mbox{EMR}^{\pi^0}}
\newcommand{\EMRiso}{\mbox{EMR}^{3/2}}
\newcommand{\CMRpi}{\mbox{CMR}^{\pi^0}}
\newcommand{\CMRiso}{\mbox{CMR}^{3/2}}
\newcommand{\tEMRpi}{\widetilde{\mbox{EMR}}^{\pi^0}}
\newcommand{\tCMRpi}{\widetilde{\mbox{CMR}}^{\pi^0}}
\newcommand{\tEMRiso}{\widetilde{\mbox{EMR}}^{3/2}}
\newcommand{\tCMRiso}{\widetilde{\mbox{CMR}}^{3/2}}
\newcommand{\tl}{\tilde\lambda}
\newcommand{\M}{M{\o}ller~}
\newcommand{\rlt}{\rho_{LT}}
\newcommand{\degree}{$^{\circ}\ $}
\newcommand{\degreeT}{$^{\circ}$}
\newcommand{\pimissmass}{$m^{\pi^{0}}_{miss} $}
\newcommand{\figref}[1]{\figurename~\ref{#1}}
\newcommand{\tabref}[1]{\tablename~\ref{#1}}

\hyphenation{Spec-tro-me-ter}

\title{Measurement of the LT-asymmetry in $\pi^0$  electroproduction
       at the energy of the $\Delta (1232)$  resonance}
\author{
D. Elsner\inst{1}
            \thanks{corresponding author, email: elsner@physik.uni-bonn.de},
A. S\"ule\inst{1}
            \thanks{email: suele@physik.uni-bonn.de},
P. Barneo\inst{2},
P. Bartsch\inst{3},
D. Baumann\inst{3},
J. Bermuth\inst{3},
R. B\"ohm\inst{3}
D. Bosnar\inst{3}
         \thanks{permanent address: Department of Physics, University of
                                    Zagreb, Croatia},
M. Ding\inst{3},
M. Distler\inst{3},
D. Drechsel\inst{3},
I. Ewald\inst{3},
J. Friedrich\inst{3},
J.M. Friedrich\inst{3}
              \thanks{present address: Physik Department E18,
                                       TU M{\"u}nchen, Germany},
S. Gr\"ozinger\inst{3}
              \thanks{present address: GSI, Darmstadt, Germany},
P. Jennewein\inst{3},
S. Kamalov\inst{3},
F. H. Klein\inst{1},
M. Kohl\inst{4}
       \thanks{present address: MIT/Bates, Massachusetts, USA},
K.W. Krygier\inst{3},
H. Merkel\inst{3},
P. Merle\inst{3},
U. M\"uller\inst{3},
R. Neuhausen\inst{3},
Th. Pospischil\inst{3},
M. Potokar\inst{5},
G. Rosner\inst{3}
         \thanks{present address: Dept. of Physics and Astronomy,
                                      University of Glasgow, UK},
H. Schmieden\inst{1},
M. Seimetz\inst{3}
       \thanks{present address: DAPNIA/SPhN, CEA Saclay, France},
O. Str\"ahle\inst{3},
L. Tiator\inst{3}
Th. Walcher\inst{3},
M. Weis\inst{3}
}                     
\institute{Physikalisches Institut,
           Rheinische Friedrich-Wilhelms-Universit\"at,
           D-53115 Bonn,
           Germany
           \and
	   NIKHEF,
	   Amsterdam,
	   The Netherlands
	   \and
           Institut f{\"u}r Kernphysik,
           Johannes Gutenberg--Universit{\"a}t,
           D-55099 Mainz,
           Germany
           \and 
           Institut f\"ur Kernphysik, 
	   Technische Universit\"at Darmstadt, 
	   D-64289 Darmstadt, 
	   Germany 
           \and
           Institut Jo\v zef Stefan,
           University of Ljubljana,
           SI-1001 Ljubljana, Slovenia }
\date{Received: date / Revised version: date}
%
\abstract{ 
The reaction \peeppi has been studied at $Q^{2}$=0.2 (GeV/c)$^{2}$ in the region
of W=1232 MeV. From measurements left and right of $\vec q$, cross section
asymmetries $\rlt$ have been obtained in forward kinematics $\rlt(\theta_{\pi^0}^{cm}= 20^{\circ}) = (-11.68 \pm 2.36_{stat} \pm 2.36_{sys})$ and backward kinematics $\rlt(\theta_{\pi^0}^{cm}=160^{\circ}) =(12.18 \pm 0.27_{stat} \pm
0.82_{sys})$ $\pi^{0}$. Multipole ratios  $\Re\{S_{1+}^{\ast}M_{1+}\} / {|M_{1+}|^{2}}$ and
$\Re\{S_{0+}^{\ast}M_{1+}\} / {|M_{1+}|^{2}}$ were determined in the
framework of the MAID2003 model. The results are in agreement with older
data. The unusally strong negative $\Re\{S_{0+}^{\ast}M_{1+}\} / {|M_{1+}|^{2}}$
required to bring also the result of Kalleicher \textit{et al.} in accordance
with the rest of the data is almost excluded.
\PACS{
      {13.60.Le}{Meson production} \and
      {13.40.-f}{Electromagnetic processes and properties} \and
      {14.20.Gk}{Baryon resonances with S=0} 
     } 
} 
\authorrunning{D. Elsner, A. S\"ule et al.}
\titlerunning{Measurement of the LT-asymmetry in $\pi^0$
              electroproduction at the energy of the $\Delta(1232)$}
\maketitle
\section{Introduction}
\label{sec:introduction}

The nucleon ground- and excited state properties pres\-ently 
elude a consistent description in terms of QCD as the basic
theory of strong interaction,
due to the non--linear, non--perturbative interaction of quarks and gluons.
Over the last years, considerable efforts aimed at a better
understanding of this complicated structure,
both theoretically and experimentally.
One important issue is the understanding of the `shape' of the
nucleon.
Despite its spin of 1/2 and, in consequence,
the vanishing spectroscopic quadrupole moment,
the nucleon wave function might have quadrupole components
which are expected to exhibit in the transition of the ground
state to the spin 3/2 $\Delta(1232)$ excitation.
Within constituent quark models those components originate from tensor forces
generated by a color hyperfine interaction
\cite{deRujula75,IKK82,GD82,DG84}.
Larger quadrupole strengths are expected from models emphasizing
the particular role of pions via exchange currents \cite{Buchmann98}
or the `pion cloud' \cite{Bermuth88,WH97,Gellas99,Silva00,Amoreira00},
and also in first quenched Lattice QCD calculations \cite{Alex04}.
Dynamical approaches \cite{KY99,KY01,SL01} enable a
decomposition into the ``bare'' contributions, as described in quark models,
and the ``dressing'' by the pion cloud. 

The quadrupole strength is usually characterized by the ratios
$R_{\rm EM} = E_{1+}/M_{1+}$ and $R_{\rm SM} = S_{1+}/M_{1+}$
of the $\pi N$ multipoles in the $\Delta(1232) \rightarrow N\pi$
decay\footnote{Exact definition and aspects of isospin separation see \cite{HS98,HS01}.},
which are uniquely related to the photon multipoles of electromagnetic
excitation \cite{DT92,Drechsel99}. Hence, these ratios can be measured in photo-
and electroproduction of pions in the energy region of the $\Delta(1232)$ resonance. 
Since unwanted non--resonant contributions are strongly suppressed
in the $\pi^0$ channel compared to the charged pion production, 
most measurements focused on the $\gamma^{(*)} p \rightarrow p \pi^0$ reaction.

A number of studies pursued at the laboratories providing cw electron
beams yielded precise coincidence data based on high luminosity beams and
high resolution detectors with large angular coverage. Partially
single or double polarization observables have been measured.
The evolution of $R_{\rm EM}$ and $R_{\rm SM}$ with negative squared
four--momentum transfer, $Q^2$, has been investigated 
over a large range in $Q^2$ up to 4\,(GeV/c)$^2$ \cite{Frolov99,Gothe00,Joo02}.
The extraction of the quadrupole ratios from the measured cross sections is
non-trivial.
For the $R_{\rm SM}$ discussed
here, it is more reliable at lower $Q^{2}$ where
$M_{1+}$ dominance is more pronounced than at higher $Q^2$
and single \cite{Warren98,Bartsch02} and double polarization results
\cite{Pospischil01,Nikhef02,CEBAF03}
are already available in addition to unpolarized recent
measurements \cite{Gothe00,Joo02,Kalleicher97,Wacker98,Mertz01,Spar04}.
The low $Q^2$ results are almost all compatible with each other,
yielding $R_{\rm SM} \simeq -6\,\%$, cf. fig. \ref{fig:cmr}.
The only exception is the result of Kalleicher \textit{et al.} \cite{Kalleicher97}.
However, due to the particular kinematics it could 
be interpreted in line with the other results,
if the ratio $S_{0+}/M_{1+} \simeq -10\,\%$ \cite{HS01}. $\snpx$ is 
related to the spin 1/2 $\rightarrow$ 1/2 transition. This amplitude was neglected in the
analysis of \cite{Kalleicher97}.
Both magnitude and sign of such an $S_{0+}$ are however unexpected from models, \textit{e.g.} MAID2003 \cite{Drechsel99}, but not excluded
by older measurements with large errors \cite{Siddle71,Alder72} which yielded
slightly positive values with errors of the order $10\,\%$ absolute.

In order to investigate this issue, measurements of
$\pi^0$ electroproduction in forward and backward direction have been
performed, which are reported in this paper. It is organized as follows:
In the next section the cross section formalism is briefly
summarized and the method is motivated. The description of the experiment is then followed by a discussion
of the data analysis, systematic error contributions and the results in sections
\ref{sec:analysis}, \ref{sec:errors} and
\ref{sec:results}.

\section{Cross section of pion electroproduction}
\label{sec:cross}

In one-photon-exchange approximation the fivefold differential cross section
of pion electroproduction
\begin{equation}
  \frac{d^5\sigma}{dE_e d\Omega_e d\Omega_{\pi}^{\rm cm}} =
  \Gamma \frac{d^2\sigma_v}{d\Omega_{\pi}^{\rm cm}}
\end{equation}
factorizes into the virtual photon flux
\begin{equation}
  \Gamma = \frac{\alpha}{2 \pi^2} \frac{E'}{E}
  \frac{k_{\gamma}}{Q^2} \frac{1}{1-\epsilon}
\end{equation}
and the virtual photon cm cross section
$d^2\sigma_v/d\Omega_{\pi}^{\rm cm}$.\linebreak[4] 
Here $\alpha$  denotes the fine structure constant,
$k_\gamma = (W^2 - m_p^2)/2m_p$  the laboratory energy of a real photon
for the excitation of the target with mass $m_p$  to the cm energy $W$, and
$\epsilon = [ 1 + ( 2 |\vec q|^2/Q^2 ) \tan^2{\frac{\vartheta_e}{2}} ]^{-1}$
the photon polarization parameter.
$Q^2 = |\vec q|^2 - \omega^2$  is the negative squared four-momentum transfer,
$\vec q$  and $\omega$  are the three-momentum and energy transfers,
respectively, and $E$, $E'$  and $\vartheta_e$
the incoming and outgoing electron energy
and the electron scattering angle in the laboratory frame.

The unpolarized cross section for pion production with virtual photons
is given by \cite{DT92,Drechsel99}
\begin{eqnarray}
  \frac{d^2\sigma_v}{d\Omega_\pi^{\rm cm}} &=:& \sigma_v = \sigma_T + \epsilon_L \sigma_L \nonumber \\ & &
  + \sqrt{2 \epsilon_L (1+\epsilon)} \sigma_{LT} \cos \phi + \epsilon \sigma_{TT} \cos 2\phi .
  \label{eq:x-sec} 
\end{eqnarray}
The partial differential cross sections, for which we use the short-hand
notation $\sigma_i$, describe the response of the hadronic
system to the polarization of the photon field, characterized by the degrees of transverse (T) and longitudinal (L)
polarization, $\epsilon$  and $\epsilon_L = \frac{Q^2}{\omega^2_{\rm cm}}
\epsilon$, respectively.
The angle $\phi$ is the tilting angle between the electron scattering plane and the reaction plane.
At $\phi=0$\degree and $180$\degree ($\phi=90$\degree and $270^{\circ}$) pions
are ejected in (perpendicularly to) the scattering plane.           
\begin{figure*}[h!t]
  \epsfig{figure=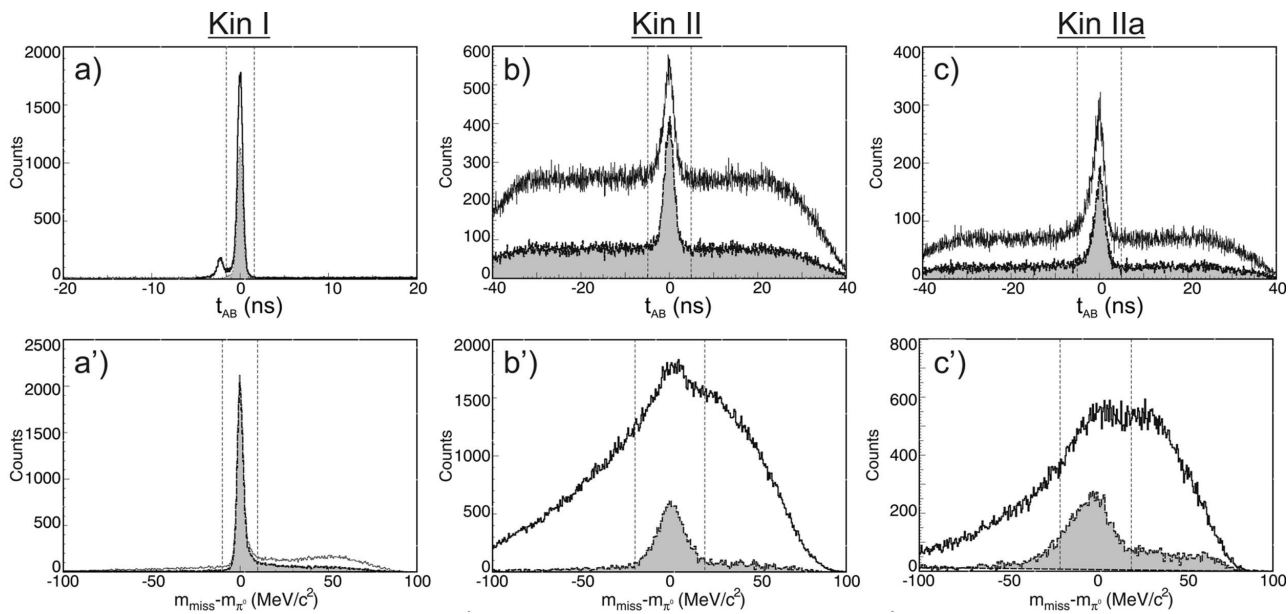 ,width=17cm}
  \caption{Typical coincidence-time (a,b,c) and missing-mass
    (a$^{\prime}$,b$^{\prime}$,c$^{\prime}$) spectra for Kin. I, II and IIa. Light spectra result from standard cuts without phase-space restrictions. 
    Shaded time spectra (FWHM peaks: (a) 0.8 ns, (b) 2.7 ns, (c) 3.0 ns)
    result from missing-mass cuts (dashed vertical lines in the missing-mass plots)
    around the $\pi^{0}$ mass; in Kin. I and Ia (not shown) the cut eleminates the $\pi^{-}$-time
    peak and at Kin. II and IIa the prompt time peak becomes symmetric (see text).
    The shaded missing-mass spectra result similarly from the indicated cuts around the coincidence-time peak.
  }
  \label{fig:1}       
\end{figure*}

\section{Method}
\label{sec:method}
The partial cross section $\sigma_{LT}$ is sensitive to both $\snpx$ and $\sepx$.
It can be determined from a fit of the $\phi$-dependence of the cross section
of eq. (\ref{eq:x-sec}). 
To this end, two measurements left ($\phi=0^\circ$) and right ($\phi=180^\circ$)
of the $\vec{q}$ direction are sufficient, which allow to form the asymmetry
\begin{eqnarray}
  \label{formel:rhoLT_phi}
  \rho_{LT}(\theta_{\pi^0}^{cm}) := 
  \frac{\sigma_{v}(\phi=0^{\circ})-\sigma_{v}(\phi=180^{\circ})}
       {\sigma_{v}(\phi=0^{\circ})+\sigma_{v}(\phi=180^{\circ})}
\end{eqnarray}
as a function of the $\pi^{0}$ center-of-mass polar angle, $\theta_{\pi^0}^{cm}$.
According to eq. (\ref{eq:x-sec}), it is related to the partial cross
sections via
\begin{eqnarray}
  \rho_{LT}(\theta_{\pi^0}^{cm}) = 
  \frac{\sqrt{2\epsilon_{L}(\epsilon+1)}\sigma_{LT}}{\sigma_{T}+\epsilon_{L}\sigma_{L}+\epsilon \sigma_{TT}}. 
  \label{formel:rhoLT}
\end{eqnarray}
The sensitivity to $\snpx$ and $\sepx$ is shown by a partial wave
decomposition of eq. (\ref{formel:rhoLT}), where only the leading
multipoles are retained. At the $\Delta (1232)$ resonance position
the asymmetry
\begin{eqnarray}
  \label{formel:rho-vor-rueck}
  \rho_{LT}(\theta_{\pi^0}^{cm}) \simeq
  f(\theta_{\pi^0}^{cm}) \cdot \frac{\Re\{(S_{0+}^{\ast} + 6  S_{1+}^{\ast} \cos\theta_{\pi^0}^{cm})M_{1+}\}}{|M_{1+}|^{2}}
\end{eqnarray}
is obtained.
Thus measurements of $\rlt$ in the
forward ($\theta_1$) and backward cm-hemisphere ($\theta_2$\nolinebreak = $\pi$ - $\theta_1$) allow the extraction of $\sepx/\mep$ and $\snpx/\mep$:
\begin{eqnarray}
  \label{formel:rho-diff}
  \frac{\Re\{S_{1+}^{\ast}M_{1+}\}}{|M_{1+}|^{2}} = f_{1}(\theta_{1,2})\cdot[\rho_{LT}(\theta_1)-\rho_{LT}(\theta_2)]+C_{1}\\
  \label{formel:rho-sum}
  \frac{\Re\{S_{0+}^{\ast}M_{1+}\}}{|M_{1+}|^{2}} = f_{0}(\theta_{1,2})\cdot[\rho_{LT}(\theta_1)+\rho_{LT}(\theta_2)]+C_{0}.
\end{eqnarray}
The functions $f_{0}(\theta_{1,2})$ and $f_{1}(\theta_{1,2})$ denote kinematical factors, $C_{0}$ and $C_{1}$ contain
contributions of multipoles beyond the approximation.

\section{Experiment}
\label{sec:experiment}
 
The \peeppi experiment was performed at the Mainz Microtron MAMI \cite{Mami90} using a beam
energy of 855 MeV and currents of $\sim33$ $\mu A$ which were measured with high
precision by a F\"orster probe in the recirculation path of the $3^{rd}$
microtron stage. The beam hit a liquid hydrogen target. Specifically designed
for this experiment, the $\varnothing$ 1 cm cylindrical target cell with
6.25 $\mu m$ Havar walls \cite{Elsner00} enabled the detection of very
low-energetic protons. The scattered electrons were detected at a central
angle of $\theta_{e^-}^{lab}$=44.45\degree and central momentum of p=408.7 MeV/c in Spectrometer A of the Three-Spectrometer
setup of the A1-collaboration \cite{SpecA1}. It consists of a QSDD magnetic
system and is equipped with two double planes of vertical drift chambers for measurement of particle
trajectories in the focal plane. During the course of the measurements presented
here, the standard Cherenkov detector for $\pi^{-}/e^{-}$-discrimination was
not available, since it was replaced by a focal-plane proton polarimeter
\cite{Pos02FP} for other experiments \cite{Pospischil01,Pos01Gep}.
In coincidence with the scattered electron, the recoil protons of the \peeppi reaction were detected  
in Spectrometer B with a similar focal-plane instrumentation. 
The smallest possible angle between Spectrometer B and the exit beam-pipe
is 9\degree and the momentum-threshold for the proton-detection is 250 MeV/c.
Hence $Q^{2}$=0.2 (GeV/c)$^{2}$ was the minimum
possible momentum transfer that could be reached at W=1232 MeV.
\begin{table}[h]
  \caption{Proton kinematics.} \label{tab:proton kinematics}
  \begin{tabular}{ccccc}
    \hline
    Kin. & $\theta_{\pi^0}^{cm}$ (\degreeT)  & $\phi$ (\degreeT)  &p$^{lab}_{p}$ (MeV/c) &
    $\theta^{lab}_{p}$ (\degreeT) \\
    \hline
    \hline
    I  & 160 & 0   & 741.7 & 33.0/32.0\\
    Ia &     & 180 &        & 20.9/21.9 \\
    \hline
    II  & 20 & 0   &  265.02 & 44.2/43.7 \\
    IIa &    & 180 &         & 9.8/10.3  \\
    \hline
  \end{tabular}
\end{table}
The four different kinematic settings are summarized in table \ref{tab:proton kinematics}.
In order to check for false asymmetries, possibly caused by inefficiencies of the focal
plane detectors in the proton arm, Spectrometer B was displaced by 1\degree
against the nominal setting for part of the measurements.\\ 
\begin{table}[h]
  \caption{Elastic scattering kinematics.}
  \label{tab:elastic kinematics}
  \begin{tabular}{cccccc}
    \hline
    \multicolumn {2}{c}{Spec. A (electron)} & \multicolumn {2}{c}{Spec. B (proton)}& Beam\\
    \hline
    \hline
    $\theta$ (\degreeT) & p (MeV/c)  & $\theta$ (\degreeT) & p (MeV/c) & E (MeV) \\
    \hline
    55.5 & 612.0 & 43.6 - 46.0 & 704.4 & 855.0 \\
    46.5 & 314.6 & 58.4, 59.4 & 251.0 & 351.3 \\
    \hline
  \end{tabular}
\end{table}
The $\pi^{0}$-data were supplemented by elastic \peep measurements (table
\ref{tab:elastic kinematics}) to monitor the
overall experimental consistency with high precision.
The overdetermined kinematics allows comparison of every measured variable with
the values calculated from the other measured variables. Corrections of 0.5 MeV/c for the central momentum of the
electron-spectrometer and 0.1\degree for the angle of the proton spectrometer
were determined.
The probable origin is a very slight mismatch between hardware (detector angle, field
integral) and the track reconstruction.

Precise measurements of electron beam current and
dead time allowed an accurate determination of the effective luminosity.

\section{Data analysis}
\label{sec:analysis}

Typical coincidence-time and missing-mass spectra are \linebreak[4] 
shown in fig. \ref{fig:1}. 
The overdetermined kinematics allows the reconstruction of the unobserved
$\pi^{0}$ by its missing mass, \pimissmass. Basic background reduction is
obtained by coincidence-time cuts and subtraction of random coincidences via
sidebands. 
In addition to the almost back\-ground-free e$^{\prime}$-p coincidence peak, the
time spectrum for the high proton-momentum kinematics shows a smaller second peak at
$\sim$ -2.2 ns (Kin. I, cf. fig. \ref{fig:1}a). It is caused by negative pions,
predominantly from $\pi^+ \pi^-$ reactions,
the $\pi^-$ of which are detected in the electron spectrometer after a longer flight time compared to electrons. 
These events can be eliminated by the coincidence-time cut indicated in fig.
\ref{fig:1}a. However, for  Kin. II and IIa, the
unwanted negative pions can no longer be separated by coicidence time, due to
insufficient time resolution caused by multiple scattering at the low proton-momentum.
Instead, additional missing-mass cuts are used to suppress these events. As also
illustrated in fig. \ref{fig:1}, with a cut around $m_{\pi^{0}}$
(a$^{\prime}$, b$^{\prime}$, c$^{\prime}$) the $\pi^{-}$ peak vanishes in
Kin. I (shaded area of fig. \ref{fig:1}a). Under the conditions of
Kin. II/IIa the resulting coincident-time peak becomes symmetric after
the missing-mass cut.

Standard cuts ensure valid track reconstruction in both spectrometers. No
target-vertex cuts were applied in order to avoid artificial $\rlt$-asymmetries
from the very different vertex-resolution along the beam-direction for the different settings.
Spectrometer acceptances were normalised with standard Monte-Carlo phase-space
simulations,\linebreak[4] which also include the radiative corrections \cite{DMW01}.

\begin{figure}
\resizebox{0.47\textwidth}{!}{%
  \includegraphics{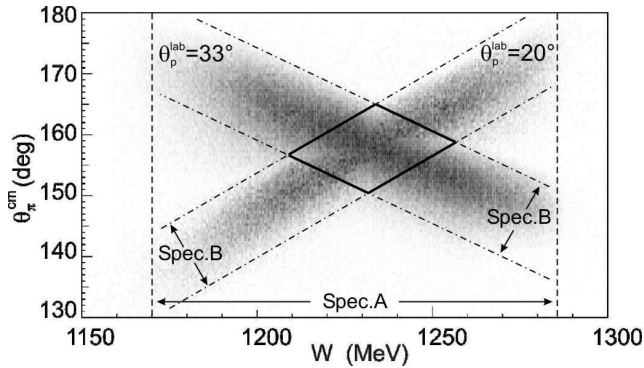}
}
\caption{Spectrometer acceptances for left ($\theta_{p}^{lab}$=33\degree) and
  right ($\theta_{p}^{lab}$=20\degree) settings of kinematics I. W is limited by the
  electron-spectrometer (A) acceptance. The angular acceptance of the proton-spectrometer (B) determines the width
  in $\theta_{\pi^0}^{cm}$.
}
\label{fig:corel} 
\end{figure}
The limited spectrometer acceptances cause different correlations between $W, Q^2, \epsilon,
\theta_{\pi^0}^{cm}$ and $\phi$ for the settings left and right of $\vec q$,
as illustrated in  fig. \ref{fig:corel}.
Due to these correlations, equal binning in the variables nevertheless leads  to
unequal distributions left and right of $\vec q$. Thus artificial
$\rlt$-asymmetries can be generated, if the mean values of the kinematic variables
differ between left and right. This is obvious, \textit{e.g.}, for a case where $W_{left}=1232MeV
- \delta$ and $W_{right}=1232MeV + \delta$, since the trivial W dependence of the
cross section produces a $\rlt\neq0$. 

It is extremely important to base the experimental asymmetries on left-right bins with the same mean
values of the variables $W, Q^{2}, \theta_{\pi^0}^{cm}$ and $\phi$. This is ensured by projection of
the numbers measured in 
each bin to the same "nominal kinematics".
For this projection we made use of the MAID2000 parametrisation.
The projection factors are obtained as the calculated ratios of
differential cross sections. In order to minimise the projection error, only data
are used within the $\theta_{\pi^0}^{cm}$--W overlap region of the two acceptance bands
in fig. \ref{fig:corel}.
Remaining uncertainties are included in the systematic error.

The appropriately normalised and projected numbers of events left (l) and right
(r) of $\vec{q}$ are determined by 
\begin{eqnarray}
  \label{formel:erg_count}
  n_{l(\phi=0^{\circ}),r(\phi=180^{\circ})} =  \frac{\sum_{Bins} (N_{l,r}/P_{l,r}) \cdot
    {\mathrm{MAID}^{corr}_{l,r}}}{L_{l,r}}.
\end{eqnarray}
Here $N_{l,r}$ denotes the number of counts after cuts, which has to be divided
by the relative phase-space acceptance $P_{l,r}$.
${\mathrm{MAID}^{corr}_{l,r}}$ is the projection factor
and $L_{l,r}$ represents the relative luminosity. The asymmetry is then simply given by
\begin{eqnarray}
  \label{formel:erg_rho}
  \rho_{LT}(\theta_{\pi^0}^{cm})  = \frac{n_{l} - n_{r}}{n_{l} + n_{r}} .
\end{eqnarray}

\section{Systematic errors}
\label{sec:errors}
The systematic error has been estimated for Kin. I/Ia from the data themselves by
variation of all kinematic cuts. For the data with low proton-momentum
(Kin. II/IIa) such an analysis is limited by the available statistics and
experimental resolution. 
The sliding cuts in the variables W and \pimissmass \ resulted in
non-negligible systematic errors (table \ref{tab:sys}a, \ref{tab:sys}b). 
The sliding cut in \pimissmass \ sets a limit on remaining radiative
effects beyond those included in the phase-space simulation.
\begin{table}[!b]
  \caption{Absolute systematic errors ($\Delta \rho_{LT}$) of high proton-momentum
    kinematics}
  \label{tab:sys}
  \begin{tabular}{lc} 
    \hline 
    a) W cut  & 0.29 \%  \\
    b) \pimissmass \  cut & 0.23 \%\\
    c) Spectrometer corrections \qquad \qquad & 0.57 \%\\
    d) MAID2000 projection  & 0.46 \%\\
    \hline{\rm } 
  \end{tabular} 
\end{table} 
The spectrometer correction,
which was determined by elastic
measurements, has been taken into account both in analysis (track
reconstruction) and simulation. The value given in table \ref{tab:sys}c 
results from the variation of the angle of Spectrometer B by $\pm 0.1^{\circ}$. 
Potential error contributions of the MAID-projection were 
estimated through relative variation of $M_{1+}$ by $\pm$5\% and, simultaneously, of $S_{1+}$ and $S_{0+}$
by $\pm$ 50\% in the full MAID2000 calculation. The largest deviation is given in table
\ref{tab:sys}d.
Additional errors for the luminosity determination are not required.
The maximum variations of 2\%-relative 
were corrected, and the remaining effect is negligible.
All kinematic settings were measured repeatedly to avoid time-dependent effects,
\textit{e.g.} efficiency variations.
These data sub-sets were combined for the left- and
right-kinematics.\\

\section{Results and discussion}
\label{sec:results}

\begin{figure} 
\resizebox{0.5\textwidth}{!}{%
  \includegraphics{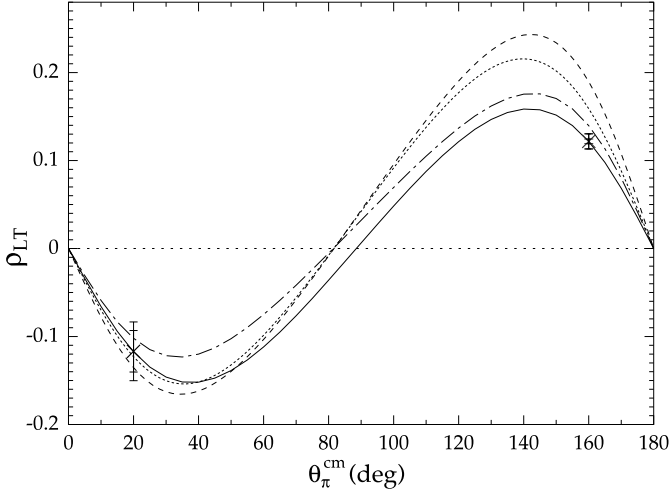} 
}
\caption{The measured $\rlt$ asymmetries compared with model predictions from MAID2003 \cite{Drechsel99} (dotted), 
  DMT2001 \cite{KY99,KY01} (dashed), Sato/Lee
  \cite{SL01} (dashed dotted). The full curve represents the MAID2003 re-fit
  reported in this paper. The depicted errors represent the statistical
  (inner bars) and the quadratic sum of 
  the statistical and systematical errors (outer bars) as discussed in the text.
}
\label{fig:erg}       
\end{figure}

\begin{figure} 
\resizebox{0.5\textwidth}{!}{%
  \includegraphics{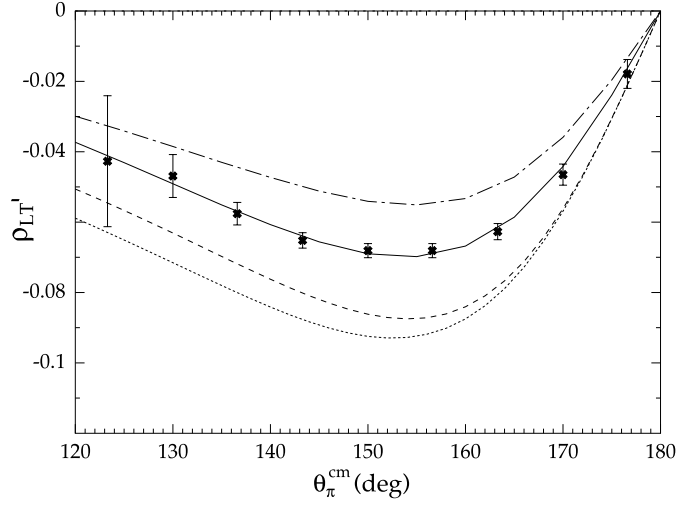} 
}
\caption{ Results for $\rlt\prime$ from reference \cite{Bartsch02} with model 
  predictions from MAID2003 \cite{Drechsel99} (dotted), 
  DMT2001 \cite{KY99,KY01} (dashed), Sato/Lee \cite{SL01} (dashed dotted). 
  The full curve represents the MAID2003 re-fit. The depicted errors are only statistical. 
}
\label{fig:erg2}       
\end{figure}

From eq. \ref{formel:erg_rho} the asymmetries
\begin{eqnarray}
  \label{formel:rhoLT_result}
  \rlt(\theta_{\pi^0}^{cm}=160^{\circ}) =(12.18 \pm 0.27_{stat} \pm 0.82_{sys})\% \nonumber \\
  \rlt(\theta_{\pi^0}^{cm}= 20^{\circ}) = (-11.68 \pm 2.36_{stat} \pm 2.36_{sys})\% \nonumber 
\end{eqnarray}
are determined.
The total systematic error 
is obtained by quadratic summation of the individual contributions in table
\ref{tab:sys}. For the forward measurement a systematic error of the same size
as its statistical one is assumed as a worst case estimate.
Using these new data in conjunction with the previous measurement of the
$\rho_{LT\prime}$ asymmetry (fifth structure function) of Bartsch et
al. \cite{Bartsch02}, we performed a re-fit of the MAID2003 parameters. We
obtained sensitivity to real and imaginary parts of the $\sepx$ and $\snpx$
amplitudes in the $p\pi^0$ channel. The results for $\rho_{LT}$ and
$\rho_{LT\prime}$ are depicted in fig. \ref{fig:erg} and \ref{fig:erg2} which,
for comparison, also shows the standard MAID2003 and the calculation within the dynamical models of Kamalov/\-Yang (DMT2001)
\cite{KY99,KY01} and Sato/ Lee \cite{SL01}.   
\begin{table}[h]
  \caption{Comparison of multipole ratios from data and calculations, as
  discussed in the text.} \label{tab:MultipolRatios}
  \begin{tabular}{lll}
    \hline
    & $\frac{\Re\{S_{1+}^{\ast}M_{1+}\}}{{|M_{1+}|^{2}}}$ (\%)&
    $\frac{\Re\{S_{0+}^{\ast}M_{1+}\}}{{|M_{1+}|^{2}}}$ (\%)\\
    \hline
    \hline
MAID2003 re-fit & -5.45$\pm$0.42$$ & 2.56$\pm$2.25$$ \\

from eqs. (\ref{formel:rho-diff},\ref{formel:rho-sum})& -4.78$\pm$0.69 & 0.56$\pm$3.89 \\

MAID2003 &  -6.65        & 7.98        \\

Sato/Lee &  -4.74         & 5.14         \\
    \hline
    \label{RatioTable}
  \end{tabular}
\end{table}

From our MAID re-fit we extract the results given in the first row in table \ref{tab:MultipolRatios}.
The denoted errors are due to the re-fit of $\sepx$ and $\snpx$ 
within the MAID2003 analysis taking into account the statistical and
systematical errors.
The model dependence of the extraction can be estimated from the truncated
multipole result given in the second row in table \ref{tab:MultipolRatios}. 
In the framework of this approximation we extract from the measured
$\rho_{LT}$ asymmetries the multipole ratios via eqs. (\ref{formel:rho-diff} and
\ref{formel:rho-sum})  with only leading terms in $\sepx$, $\snpx$ and $\mep$.
The last two lines in table \ref{tab:MultipolRatios}
contain standard model values without re-fit to our data.
\begin{figure}[h] 
\resizebox{0.48\textwidth}{!}{%
  \includegraphics{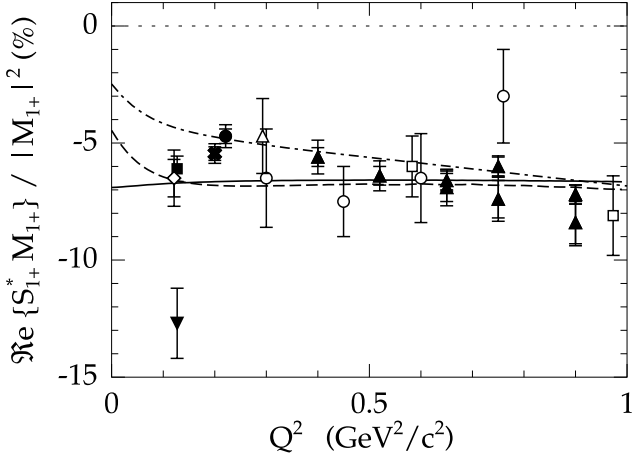} 
}
\caption{Result for $\Re\{S_{1+}^{\ast}M_{1+}\} / {|M_{1+}|^{2}}$ with statistical and
  systematical errors as extracted from this experiment using the MAID2003 re-fit (full cross), compared to measurements.
  Data where only statistical errors are given: DESY
  \cite{Siddle71} (open square), NINA \cite{Alder72} (open circles), Bonn
  synchrotron \cite{Baetz74} (open triangle tip up) and ELSA \cite{Kalleicher97}
  (full triangle tip down). Data, where statistical and systematical errors are given: ELSA \cite{Wacker98} (full circle, to improve the presentation shiftet from  $Q^{2}$=0.201
  (GeV/c)$^{2}$ to $Q^{2}$=0.221 (GeV/c)$^{2}$), MAMI \cite{Pospischil01} (open
  diamond), CLAS \cite{Joo02} (full triangles) and BATES \cite{Mertz01,Spar04}
  (full square). The curves show model calculations MAID2003 \cite{Drechsel99} (solid), DMT2001
  \cite{KY01} (dashed) and Sato/Lee \cite{SL01} (dashed dotted).
 }
\label{fig:cmr} 
\end{figure}

\begin{figure}[h] 
\resizebox{0.5\textwidth}{!}{%
  \includegraphics{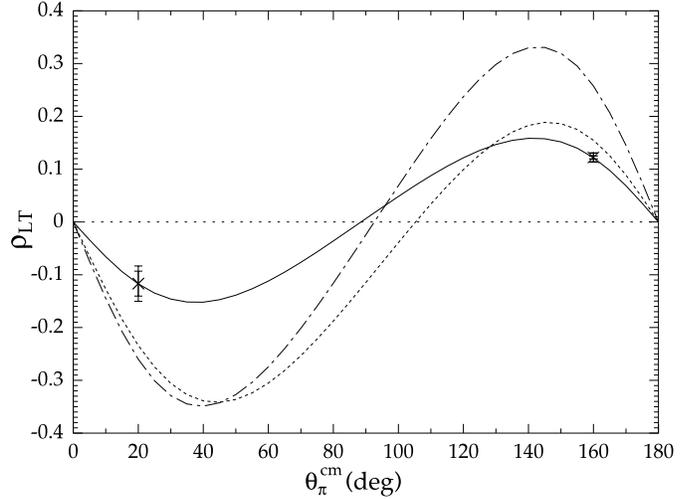} 
}
\caption{
  The plot shows, at smaller scale, the MAID2003 re-fit (solid) in comparison to
  the full MAID2003 calculation with modified $\Im\{\sepx\}$ and $\Im\{\snpx\}$ for two situations:
  $\sepx/\mep=-12.5\%$, $\snpx/\mep=0.0\%$ (dashed dottet),
  i.e. the result of \cite{Kalleicher97}, and $\sepx/\mep=-10\%$,
  $\snpx/\mep=-14\%$ (dottet), which would bring \cite{Kalleicher97} in accordance with
  other measurements \cite{HS01}.
}
\label{fig:erg3}       
\end{figure}

Figure \ref{fig:cmr} shows our full MAID2003 result for $R_{\rm SM}$.  
Within the errors, our extracted value is in accordance with measurements at the same $Q^2$ \cite{Wacker98} and measurements
at adjacent $Q^2$ \cite{Joo02,Pospischil01}.
The negative-slope tendency of the CLAS-data \cite{Joo02} seems to be further supported
by our $R_{\rm SM}$ value at smaller $Q^2$. 
Provided there is no sharp $Q^2$-dependence in $R_{\rm SM}$, we can rule out the result of Kalleicher \textit{et al.}
\cite{Kalleicher97} at $Q^2$=0.127 (GeV/c)$^{2}$. 
This has been argued before from measurements in backward kinematics
\cite{Mertz01} where an $S_{0+}$ contribution can not be excluded. In contrast,
our conclusion comes from forward kinematics as also exploited by \cite{Kalleicher97}. 
This is illustrated in in fig. \ref{fig:erg3}. On a different scale it shows the same $\rlt$-asymmetry as
fig. \ref{fig:erg}, the Maid2003 re-fit and a full  MAID2003 calculation
using the multipole ratios $S_{0+}/M_{1+}$ and $S_{1+}/M_{1+}$ of Kalleicher \textit{et al.}. 
In addition, a full MAID2003 calculation using $S_{0+}/M_{1+} \simeq -10\,\%$, 
which could reconcile the Kalleicher result with others \cite{HS01}, 
is shown. However, this is clearly excluded by our measurement at
$\theta_{\pi^0}^{cm}=20^{\circ}$ with a $4 \sigma$ significance. 

\begin{figure}[ht] 
  \resizebox{0.48\textwidth}{!}{%
    \includegraphics{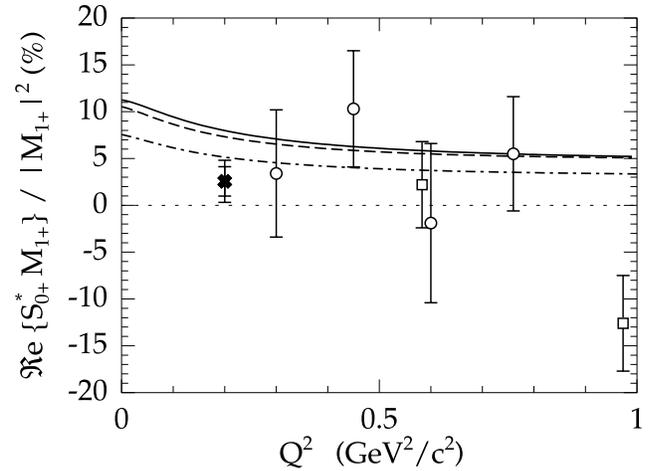} 
  }
\caption{Result for $\Re\{S_{0+}^{\ast}M_{1+}\} / {|M_{1+}|^{2}}$  with statistical and
  systematical errors as extracted from this experiment using the MAID2003 re-fit (full
 cross); compared to measurements from DESY
  \cite{Siddle71} (open square), NINA \cite{Alder72} (open circles), where
  only statistical errors are indicated. 
  The curves show model calculations MAID2003 \cite{Drechsel99} (solid), DMT2001
  \cite{KY01} (dashed) and Sato/Lee \cite{SL01} (dashed dotted).
}
\label{fig:s0+} 
\end{figure}

Our $S_{0+}/M_{1+}$ ratio, extracted from 
the MAID2003 re-fit, is plotted in fig. \ref{fig:s0+}.Within the errors it
agrees with older data at slightly larger $Q^2$ \cite{Siddle71,Alder72}.
Although a little lower, it is also compatible with the Sato/Lee, DMT2001 and standard MAID2003
parametrisations. In view of the quite large experimental errors it is not yet 
clear whether this ratio differs from zero. But we can not support a large negative $S_{0+}/M_{1+}$ ratio.

A sensitive access to the ratio $\snpx$/$\sepx$ is provided by a precise measurement of the zero-crossing
of $\rho_{LT}$ or $\sigma_{LT}$.
By now it is possible to extract this ratio from available data close to the
zero crossing at $Q^2$=0.127 (GeV/c)$^{2}$ \cite{Spar04}, with the result compatible to the MAID2003 parametrisation.  Other existing data \cite{Joo02} cover the full range of $\theta_{\pi^0}^{cm}$ at
$Q^2$=0.4-1.8 (GeV/c)$^{2}$. However, at higher $Q^2$ the $\snpx$ extraction
seems to be affected more strongly by higher partial waves than expected in the paper by
Joo \textit{et al.} \cite{Joo02}. This might be resolved by very recent
polarisation data \cite{Kell05a,Kell05b}.\\
In future experiments at MAMI-C a more accurate determination of $\snpx$
at low $Q^2$ is feasible, using the Three-Spectrometer setup of the
A1-collaboration complemented by the KaoS-spectrometer \cite{KaoS}.

\section{Summary}
\label{sec:summary}
We have measured the $\rho_{LT}$ asymmetry
in forward \linebreak[4]($\theta_{\pi^0}^{cm}$=20\degreeT) and backward
($\theta_{\pi^0}^{cm}$=160\degreeT) kinematics of $\pi^{0}$ electroproduction off
the proton at $Q^{2}$=0.2 (GeV/c)$^{2}$ \linebreak[4]around W=1232 MeV.
The measurement of the two kinematic settings allows the extraction of 
$\sepx$ and $\snpx$ in a very transparent way within a simple s- and p-wave approximation or,
alternatively, using the full MAID2003 parametrisation without any truncation.
Our results for $S_{1+}/M_{1+}$ and $S_{0+}/M_{1+}$ are in agreement with
existing measurements and calculations. Our result removes a remaining possibility
to reconcile the datum of Kalleicher \textit{et al.} \cite{Kalleicher97} for the ratio
$S_{1+}/M_{1+}$ with other measurements through a large negative  $S_{0+}/M_{1+}$.\\
We thank T.-S.H. Lee and T.Sato for providing their calculations.
This work was supported by the Deutsche \linebreak[4]Forschungsgemeinschaft (SFB 443).
%

\end{document}